\documentstyle[12pt]{article}

\begin{document}
\author{J.~W.~Bos$^1$, D.~Chang$^{2,1}$, S.~C.~Lee$^1$, Y.~C.~Lin$^2$,\\
  and H.~H.~Shih$^3$
  \\
  {\small ${}^1$Institute of Physics, Academia Sinica, Taipei, Taiwan
    }
  \\
  {\small ${}^2$Department of Physics, National Tsing Hua University,
    Hsinchu, Taiwan }
  \\
  {\small ${}^3$Department of Physics and Astronomy, National Central
    University, Chungli,}\\{\small Taiwan} }

\title{$SU(3)$ breaking and baryon magnetic moments}

\date{}

\maketitle

\begin{abstract} 
  We show that the magnetic moments of the octet baryons can be fitted
  to an accuracy of 1.5 \%
  by a phenomenological Lagrangian in which $SU(3)$
  breaking corrections appear only linearly.  This is in contrast
  to conventional chiral perturbation theory in which corrections
  non-analytic in $SU(3)$ breaking dominate and tend to spoil the
  agreement with the data.  Motivated by this observation, we propose
  a modified scheme for chiral perturbation theory that gives rise to
  a similar linear breaking of $SU(3)$ symmetry.
  (Pacs 13.40.Em, 14.20.-c, 11.30.Rd) (hep-ph/9602251)
\end{abstract}

The magnetic moments of the octet baryons were found to obey
approximate $SU(3)$ symmetry a long time ago by Coleman and Glashow
\cite{coleman}.  In the $SU(3)$ symmetric limit, the nine observable
moments (including the transitional moment between $\Sigma^0$ and
$\Lambda$) can be parameterized in terms of two parameters, and as a
result obey approximate relationships.  As we will show later, the
two parameter result of Coleman and Glashow can in fact fit the
observed magnetic moments up to about the $20 \%$ level.  However,
since at present the moments have been measured with an accuracy of
better than $1 \%$ \cite{pdb}, an improved theoretical understanding
is clearly desirable.  It is the goal of this paper to show how this can 
be achieved easily phenomenologically and in a scheme of chiral 
perturbation theory.

Many attempts were made trying to improve the numerical predictions of
Coleman and Glashow by including the $SU(3)$ breaking effects using
chiral perturbation theory (ChPT) \cite{caldi,krause,jenk,luty}.
However, many of these efforts resulted in numerical fits {\em worse\/}
than the leading order $SU(3)$ invariant one by Coleman and Glashow.
For example, Caldi and Pagels \cite{caldi} found that the leading
$SU(3)$ breaking corrections, in their scheme for ChPT, appear in the
non-analytic forms of $\sqrt{m_s}$ and $m_s \ln m_s$.  They showed that
the $\sqrt{m_s}$ corrections are in fact at least as large as the
$SU(3)$ invariant zeroth order terms, which casts doubt on the
applicability of ChPT.  Caldi and Pagels suggested that this
``failure'' of ChPT might be attributed to the large mass of kaon in
the loops and the fact that the leading correction is of non-analytic
form.  Such non-analytic contributions were indeed pointed out earlier
by Li and Pagels \cite{li} and others \cite{langacker}, however, the
non-analyticity appears only in the $SU(3)$ invariant chiral symmetry
breaking mass, not in $SU(3)$ breaking parameters.  More recently,
similar large corrections to the baryon magnetic moments non-analytic
in $m_s$ have been found, by calculating them up to the one-loop level
in ChPT \cite{krause,jenk,luty}.  By only using the $\sqrt{m_s}$ terms
Jenkins {\em et al.} \cite{jenk} could improve the accuracy of the
Coleman-Glashow results from 20 \% to about 10 \%. However, this could
only be achieved by using in kaon loops a {\em different\/} value
of the meson decay constant than in pion loops, with the effect that
the magnitude of the kaon loops is artificially reduced. In addition,
Krause \cite{krause} showed that $m_s \ln m_s$ corrections are just as
important, which disagrees with Refs.~\cite{jenk,luty}.  Also, Krause
further argued that the non-analytic contributions are not a good
approximation of the loop integrals at all. 


In the light of the problems of ChPT in accounting for the magnetic
moments, we reconsider in this paper Okubo's extension \cite{okubo} of
the phenomenological approach of Coleman and Glashow for fitting the
current accurate magnetic moment data.  Motivated by our successful
numerical results we discuss how ChPT can be formulated to avoid
corrections non-analytic in $m_s$.

Following Ref.~\cite{okubo}, we simply assume that the operators which
give the leading $SU(3)$ breaking corrections to the magnetic moments
have the same chiral transformation property as the strange quark mass
operator, and demand that the coefficients of these operators should
be of the order of $m_s / \Lambda_\chi$ ($m_s$ is strange quark mass,
and $\Lambda_{\chi}$ is chiral symmetry breaking scale).

To zeroth order in $m_s/\Lambda_\chi$ and to first order in
electromagnetic coupling, the $SU(3)$ invariant terms that contribute
to the magnetic moment
can be written in a simple form as
\begin{equation}
\label{b}
        b_1{\rm Tr}\Bigl[\,\bar{H} \hat{O} \{Q,H\}\,\Bigr]
        + b_2{\rm Tr}\Bigl[\,\bar{H} \hat{O} [Q,H]\,\Bigr]\;,
\end{equation}
where the operator $\hat{O}$ is defined as $\hat{O} =
F_{\mu\nu}\sigma^{\mu\nu}$, $H$ is the usual representation for the
baryon-octet in flavor-space given by Ref.~\cite{we}, and $Q$ is the
quark charge-matrix
\begin{equation}
        Q=\frac{1}{3}{\rm diag}(2,-1,-1)\;.
\end{equation}
The leading-order results of Coleman and Glashow are based on the
above form of the electromagnetic interaction.  As shown in Table~I,
with $b_1$ and $b_2$ one can already account for the observed baryon
magnetic moments at about the $20 \%$ level.  To analyze the magnetic
moments with the leading (linear) order of $SU(3)$ breaking
corrections included, one needs to pay attention only to the flavor
structure of the operators.  Since the operator $\hat{O}$ is flavor
independent, we will suppress it in the following.

We will assume, analogously as Okubo \cite{okubo}, that the lowest
order $SU(3)$ breaking corrections are parameterized by baryon
operators that have the same chiral transformation property as the
quark-level strange quark mass operator.  The operators in flavor
space that are of first (linear) order in $SU(3)$ breaking can be
constructed using the matrix  $m_s \sigma$, where $\sigma={\rm
  diag}(0,0,1)$.  They include four single-trace terms
\begin{eqnarray}
\label{alpha}
        &&\alpha_1 {\rm Tr}\Bigl[\,\bar{H}  [[Q,H], \sigma]\,\Bigr]
        + \alpha_2 {\rm Tr}\Bigl[\,\bar{H}  \{[Q,H], \sigma\}\,\Bigr]
        \nonumber
\\
        &&\mbox{}
        + \alpha_3 {\rm Tr}\Bigl[\,\bar{H} [\{Q,H\}, \sigma]\,\Bigr]
        + \alpha_4 {\rm Tr}\Bigl[\,\bar{H}
        \{\{Q,H\}, \sigma\}\,\Bigr]\;,
\end{eqnarray}
and four double-trace terms
\begin{eqnarray}
\label{beta}
        &&\beta_1 {\rm Tr}\Bigl[\,\bar{H} H \Bigr]\,
        {\rm Tr}\Bigl[\,\sigma Q \Bigr]\,
        + \beta_2 {\rm Tr}\Bigl[\,\bar{H} [Q,H] \Bigr]\,
        {\rm Tr}\Bigl[\,\sigma \Bigr]\,
        \nonumber
\\
        \nonumber&&\mbox{}
        + \beta_3 {\rm Tr}\Bigl[\,\bar{H} \{Q,H\} \Bigr]\,
        {\rm Tr}\Bigl[\,\sigma \Bigr]\,
\\
        &&\mbox{}
        + \Bigl(\beta_4 {\rm Tr}\Bigl[\,\bar{H} Q \Bigr]\,
        {\rm Tr}\Bigl[\,\sigma H \Bigr] + {\rm h.c.}\Bigr) \;.
\end{eqnarray}
The term with $\beta_1$ is diagonal in baryon space, i.e., contributes
equally to any baryon magnetic moment, but not to the $\Sigma
\rightarrow \Lambda$ transitional moment.  Operators associated with
$\beta_2$ and $\beta_3$ will not change the predictions based on $b_1$
and $b_2$, and can be ignored here.  The operator associated with
${\rm Re}(\beta_4)$ can, by Cayley's theorem, (see Ref.~\cite{we}) be
related to the other operators in Eqs.~(4) and~(5), while the operator
associated with ${\rm Im}(\beta_4)$ is time reversal non-invariant and
should be negligible in our analysis.  Therefore, we only have to
include $\alpha_1$, $\alpha_2$ $\alpha_3$, $\alpha_4$, and $\beta_1$
in our analysis.
 
Similar results can be found in the context of ChPT at tree-level.
For example, Eqs.~(\ref{b}--\ref{beta}) also appear as counterterms in
the chiral Lagrangians of Refs.~\cite{krause,jenk}.  However, in
contrast to ChPT in Refs.~\cite{krause,jenk}, in our approach the
flavor structure of Eqs.~(\ref{b}--\ref{beta}) is assumed to remain
valid to {\em all\/} orders in momentum expansion.

We fit the eight available magnetic moment data with the seven
parameters while requiring the resulting $\alpha_i$, and $\beta_1$ to
be a factor of order ${m_s / \Lambda_\chi}$ smaller than $b_i$ for
consistency.  The resulting magnetic moments of baryons are given in
the Table \ref{t1}.  Note that the sign of the transition moment
between $\Sigma^0$ and $\Lambda$ is a matter of convention.  The
average deviations of the fitted to the observed moments is 1.5~\%, as
compared to 20~\% for the leading order fit.

At the first sight, a seven parameter fit of eight observables may not
sound as much of an achievement.  However, the fact that the typical
values for $\alpha_i$ and the two parameters $\beta_i$ turn out to be
smaller than $b_i$ by roughly a factor of $m_s/\Lambda_\chi$ is quite
nontrivial and should be consider an evidence that the $SU(3)$
breaking effect should appear linearly in any model for magnetic
moments of baryons.  It is in sharp contrast with the ChPT in
Refs.~\cite{krause,jenk} which yield corrections which are
non-analytic in $m_s$ and are of the same size of, or larger than, the
leading-order ($SU(3)$ invariant) terms.

In addition, our result also predicts the not yet measured magnetic
moment of $\Sigma^0$ to great accuracy.  Based upon our fit, we expect
$\mu_{\Sigma^0}/\mu_N = 0.66 \pm 0.03$ (where $\mu_N$ is the standard
nuclear magneton) compared to $0.54 \pm 0.09$ from the lattice
calculation of Leinweber, Woloshyn and Draper \cite{leinweber}.  The
magnetic moment of $\Sigma^0$ may continue to be difficult to measure
directly, however, knowing its value accurately can be very useful
both theoretically and experimentally elsewhere, such as weak
radiative decays of hyperons.

Another outstanding issue \cite{lach} in the understanding of
magnetic moments is the difference between the $p$ and $\Sigma^+$
magnetic moment (which is vanishing in the $SU(3)$ symmetric limit).
The fact that this difference can be accounted for satisfactorily by
our current scheme is another support for linear $SU(3)$ breaking that
we assumed.

The assumption that $SU(3)$ is broken linearly leads to
the Okubo relations between the magnetic moments \cite{okubo}
\begin{equation}
\label{okubo1}
        \mu_{\Sigma^0-\Lambda} = \frac{1}{2\sqrt{3}}\left(
        \mu_{\Sigma^0}+3\mu_{\Lambda}-2\mu_{\Xi^0}-2\mu_n\right)\;,
\end{equation}
and
\begin{equation}
\label{okubo2}
        \mu_{\Sigma^0} = \frac{1}{2}\left( \mu_{\Sigma^+} + \mu_{\Sigma^-}
        \right)\;,
\end{equation}
where $\mu_{\Sigma^0-\Lambda}$ is the transition moment between
$\Sigma^0$ and $\Lambda$. The first relation agrees well with
experiment, while the second can not be verified yet. 


To put our result in proper perspective, one should be reminded that
so far none of the existing models, e.g. such as those derived from the
quark-model \cite{quark}, have been able to reproduce the high
precision data on the magnetic moments. On the other hand, none of the
models have employed seven free parameters.  For example, while
Ref.~\cite{jenk} may not have accounted for the magnetic moment data
successfully, the paper employs only the two $SU(3)$ invariant
parameters $b_1$ and $b_2$.  In their approach the $SU(3)$ breaking is
only due to the meson masses in the loops.  Our result should be taken
as a strong indication that we need a scheme or model in which the
leading $SU(3)$ breaking effects appears as the linear correction of
order ${m_s / \Lambda_\chi}$.

In the following we will discuss how one can modify the scheme of a
ChPT so that a polynomial expansion in $m_s$ is a result.  One way to
interpret the success of our phenomenological analysis is that the
non-analytic contributions in $m_s$ are actually not present, even in
the context of ChPT.  Such a scheme was already pointed out in the early
works on ChPT, but it was abandoned in later applications.  In the
early papers of Li and Pagels \cite{li} and Langacker and Pagels
\cite{langacker} the starting point to study deviations from chiral
symmetry was the chiral symmetry breaking Hamiltonian
\begin{equation}
\label{hamil}
        \epsilon H = \epsilon_0 H_0 + \epsilon_8 H_8\;,
\end{equation}
with the Hamiltonian $H_0$ invariant under $SU(3)$ but not under
$SU(3)\times SU(3)$, and the Hamiltonian $H_8$ breaking both $SU(3)$
and $SU(3)\times SU(3)$.  As the relevant parameters to measure the
$SU(3)\times SU(3)$ and $SU(3)$ breaking
\begin{equation}
\label{nul}
        \epsilon_0 \propto M^2_{\pi}/\Lambda^2_\chi
\end{equation}
and
\begin{equation}
\label{acht}
        \epsilon_8 \propto \left(M_K^2 - M_\pi^2\right)/
        \Lambda^2_\chi
\end{equation}
were proposed \cite{langacker}.  Note that the parameter $\epsilon_8$
is proportional to $m_s$ (assuming for simplicity that $m_u,m_d <<
m_s$).  These authors showed that infra-red divergences caused by
Goldstone bosons in the chiral loops will give rise to non-analytic
terms in $\epsilon_0$.  However, they also pointed out that an
expansion in the $SU(3)$ breaking parameter $\epsilon_8$, still
remains possible.

In ChPT this can be realized \cite{new} by taking a $SU(3)$
invariant infra-red cut-off, denoted by $M$, for the
pseudo-scalar mesons in the
loops. In this scheme all the octet mesons have the
propagator
\begin{equation}
        S(k) = \frac{i}{k^2-M^2}\;.
\end{equation}
Effects due to breaking of $SU(3)$ can then be consistently treated
perturbatively.  When the scheme is applied to the magnetic moments, it
means that all the $SU(3)$ breaking effects will remain polynomial in
$\epsilon_8$ (or equivalently $m_s/\Lambda_\chi$) and the
experimentally favored symmetry breaking pattern given by
Eqs.~(\ref{b}), (\ref{alpha}) and~(\ref{beta}) remains valid to {\em
all orders\/} in the loop expansion.  Because $M$ appears in the
denominator of the meson propagators, loop diagrams will still give
contributions non-analytic in $M$. However, since loop contributions
will have the same symmetry structure as those
Eqs.~(\ref{b}), (\ref{alpha}) and~(\ref{beta}) up to linear order in
$m_s$, these non-analytic terms can be absorbed in the coefficients
$b_i$, $\alpha_i$, $\beta_i$.  Our phenomenological analysis of the
magnetic moments will be a natural consequence of this revised scheme
of ChPT to linear order in ${m_s / \Lambda_\chi}$.

To conclude, we showed that the experimental data for baryon
magnetic moments seem to support a scheme of ChPT in which the
$SU(3)$ breaking effect can be written as a polynomial expansion in
$m_s$.  This new scheme clearly differs from the schemes of ChPT used
in Refs.~\cite{krause,jenk,luty} for the baryon magnetic moments.
In those approaches both the pion mass and the kaon mass are used as
the infra-red cut-off for their respective propagators in the loops.
This necessarily leads to $SU(3)$ breaking effects different from that
in Eqs.~(\ref{b}), (\ref{alpha}) and~(\ref{beta}) and furthermore to
terms non-analytic in $m_s$.  Since the interaction vertices
are expanded perturbatively in $m_s$, it is not consistent to use $m_s$
as propagator mass in the loops.  In addition, our analysis  shows that
the contributions non-analytic in $m_s$ are not needed to get a
satisfactory fit with the data.

This work was supported by the National Science Council of the R.O.C.
under Contracts Nos.  NSC85-2112-M-007-032, NSC85-2112-M-007-029, and
NSC85-2112-M-001-026.

%
%

\onecolumn

{\small
\begin{table}
\label{t1}
\centerline{
\begin{tabular}{c|cccc}
$B$ & ${\cal M}_{{\rm th}}/e$ & ${\cal M}_{{\rm exp}}/\mu_N$ 
& ${\cal M}_{{\rm SU(3) inv}}/\mu_N$  &
${\cal M}_{{\rm SU(3) breaking}}/\mu_N$
\\
\hline
$p$ & $b_1/3+b_2+\alpha_1+\alpha_2+(\alpha_3+\alpha_4)/3-\beta_1/3$ &
$2.793 \pm 0.000$ & 2.29 & 2.793
\\
$n$ & $ -2b_1/3-2(\alpha_3+\alpha_4)/3-\beta_1/3$ & 
$-1.91 \pm 0.000$ & -1.48 & -1.969 
\\
$\Lambda$ & $-b_1/3-8\alpha_4/9-\beta_1/3$ & 
$-0.613 \pm 0.004$ & -0.74 & -0.604
\\
$\Sigma^+$ & $b_1/3+b_2-\beta_1/3$ & 
$2.458 \pm 0.010$ & 2.28 & 2.481
\\
$\Sigma^0$ & $b_1/3-\beta_1/3$ & 
--- & 0.74 & 0.66
\\
$\Sigma^-$ & $b_1/3-b_2-\beta_1/3$ & 
$-1.160 \pm 0.025$ & -0.80 & -1.155
\\
$\Xi^0$ & $-2b_1/3+2(\alpha_3-\alpha_4)/3-\beta_1/3$ & 
$-1.250 \pm 0.014$ & -1.48 & -1.274
\\
$\Xi^-$ & $b_1/3-b_2+\alpha_1-\alpha_2-(\alpha_3-\alpha_4)/3-\beta_1/3$ &
$-0.6507 \pm 0.0025$ & -0.80 & -0.6507
\\
$\Sigma^0-\Lambda$ & $b_1/\sqrt{3}$ & 
$\pm 1.61 \pm 0.08$ & 1.28 & 1.541
\\
\end{tabular}}
\caption{Magnetic moments of the octet baryons, and transition moment
for $\Sigma^0 \rightarrow \Lambda + \gamma$, with linear $SU(3)$
breaking corrections.  The first column contains our predictions, with
the constants $b_i$ from Eq.~(\protect\ref{b}), and the constants
$\alpha_i, \beta_1$ from Eqs.~(\protect\ref{alpha})
and~(\protect\ref{beta}).  The second column contains the experimental
values  from Ref.~\protect\cite{pdb}, and the third  the
fitted values in the $SU(3)$ symmetry limit.  Finally, the fourth
column contains the fitted values based on our symmetry prediction in
column one.  In units of GeV$^{-1}$ we find $b_1=1.42$, $b_2=0.97$,
$\alpha_1= 0.32$, $\alpha_2=-0.14$, $\alpha_3=0.28$, $\alpha_4=-0.31$,
and $\beta_1 = 0.36$.  The average deviation of the predicted values
from the observed is 1.5~\%, while in the $SU(3)$ symmetric fit (see
column three) this deviation is 20~\%. We didn't weight the deviation of
each fit value from the experimental value by the experimental standard
deviation since the expected theoretical error is generally much larger
than the experimental error bar.}
\end{table}}

\end{document}